\documentstyle[11pt]{article}
\def\th{\theta}

\def\g{\gamma} \def\G{\Gamma}

\def\d{\delta} \def\D{\Delta}
\def\e{\epsilon}

\def\r{\rho}
\def\s{\sigma} \def\S{\Sigma}

\def\gkn{G_K(N)}
\def\gk{G(K)}
\def\gn{G(N)}
\def\cm{{\cal M}}
\def\be{\begin{equation}}
\def\ee{\end{equation}}
\def\loon{\relbar\joinrel\relbar\joinrel\relbar\joinrel\rightarrow}

\renewcommand{\baselinestretch}{1.0}

\begin{document}

\begin{flushright}
hep-th/9510064\\
BRX--TH--383\\
US-FT-25-95
\end{flushright}

\renewcommand{\baselinestretch}{2.0}
\small
\normalsize

\begin{center}
{\Large\bf BF Theories and Group-Level Duality}

Jos\'{e} M. Isidro\footnote{Supported by Ministerio de Educaci\'{o}n y
Ciencia (Spain).}\\
Jo\~{a}o P. Nunes\footnote{Supported by FLAD (Portugal).}\\
and \\
Howard J. Schnitzer\footnote{Research supported in part by the DOE under
grant DE--FG02--92ER40706.}\\
Department of Physics\footnote{email address: ISIDRO, NUNES, SCHNITZER@BINAH
.CC.BRANDEIS.EDU}\\
Brandeis University, Waltham, MA 02254\\
September 1995
\end{center}

\begin{center}
{\bf Abstract}
\end{center}

\renewcommand{\baselinestretch}{1.5}
\small
\normalsize

\begin{quotation}
It is known that the partition function and correlators of the two-dimensional
 topological field theory $\gkn / \gkn$ on the Riemann surface
$\S_{g,s}$ is given by Verlinde numbers, dim($V_{g,s,K}$) and that the large
$K$ limit of dim($V_{g,s,K}$) gives Vol(${\cal M}_s$), the volume of the moduli
space of flat connections of gauge group $\gn$ on $\S_{g,s}$, up to a power
of $K$.  Given this relationship, we complete the computation of
Vol(${\cal M}_s$) using only algebraic results from conformal field theory.

The group-level duality of $\gkn$ is used to show that if $\gn$ is a classical
 group, then
$\displaystyle \lim_{N\rightarrow \infty} \gkn / \gkn$ is a BF theory with
gauge group $\gk$.  Therefore this limit computes Vol(${\cal M}^\prime_s$),
the volume of the moduli space of flat connections of gauge group $\gk$.

\end{quotation}

\newpage
\renewcommand{\baselinestretch}{2.0}
\small
\normalsize

\renewcommand{\theequation}{1.\arabic{equation}}
\setcounter{equation}{0}

\noindent{\bf I. Introduction}

Two-dimensional topological theories are of considerable interest, as they
provide significant insights into the behavior of non-perturbative solutions
of quantum field theories.  An important class of topological theories
formulated on Riemann surfaces $\sum_g$ (of genus $g$) can be described as
cosets $G_K/G_K$ where $G_K$ is the affine Lie algebra at level $K$
associated to the Lie algebra of $G$, which we'll take to be compact and
semisimple.  It is known that these theories are
equivalent to another class of topological theories, the so-called BF
theories.  It can be shown that \cite{001,002}
\be 
\lim_{K\rightarrow \infty} Z_{\S{_g}} (G_K / G_K ) = \int [dA][dB] \exp  i
\{ S_{BF} (A,B)\}
\ee
where
$$
S_{BF} (A,B) = \frac{1}{2\pi} \int_{\S_g} {\rm tr} (BF_A)
$$
and
\be 
F_A = d_A \: A
\ee
In (1.1) and (1.2) $A$ is the gauge connection ($F_A$ the field strength)
appropriate to the gauge group $G(N)$, and $B$ is a scalar field in the
adjoint representation of $G(N)$. In the
large $K$ limit, (1.1) then describes the zero-area limit of (pure)
two-dimensional Yang--Mills theory with gauge group $G$.  In fact, the string
action for the two-dimensional QCD string discussed by Cordes, {\it et al.}
\cite{003}, begins with YM$_2$ in the zero-area limit.  Therefore
(1.1) provides a {\it formal} computation of Vol(${\cal M}$), the
volume of the moduli space of flat connections ${\cal M}$ of the gauge group
$G(N)$ on $\S_g$.  That is, integrating over the $B$ field,
\be 
\int [dA] \d (F_A) = {\rm Vol}(\cm )
\ee
It is significant for our purposes that Vol($\cm$) can be computed from the
Verlinde numbers \cite{005} $\dim V_{g,o,K}$, where $V_{g,o,K}$ denotes the
space of conformal blocks of $G_K(N)$ on $\S_g$.  The relevant relation is
\cite{001,004}
\be 
{\rm Vol}({\cal M}) = \lim_{K\rightarrow \infty}
\left[ K^{-\frac{1}{2} \dim  \cm} \cdot \dim V_{g,K} \right]
\ee
with
\be 
\dim  \cm = 2 (g-1) \dim G
\ee
and
\be 
\dim V_{g,K} = \sum_a (S_{oa})^{2-2g}
\ee
where $S_{ab}$ is the modular transformation matrix of $G_K(N)$, and the sum
in (1.6) is over all integrable representations of $G_K(N)$, (with $o$
denoting the identity representation).  A relation similar to (1.4) to (1.6)
holds for $s$-punctured Riemann surfaces and $\dim \cm_s$ as well, where $\cm
_s$ is the space of flat connections on the $s$-punctured Riemann surface
$\S_{g,s}$. This will be discussed in subsequent sections.

Witten \cite{001,004} has computed Vol($\cm$) and Vol($\cm_s$) using geometric
and topological techniques.
However, since the Verlinde
numbers \cite{005} $\dim V_{g,s,K}$ are expressed entirely in terms of the
modular
transformation matrices $S_{ab}$ of conformal field theory, one suspects that
{\it given} (1.4)--(1.6), and its generalization to $\sum_{g,s}$, one
should be able to compute ${\rm Vol} (\cm)$ and ${\rm Vol} (\cm_s)$ by
algebraic
methods.  This in fact is one of the results
presented in this paper.

Since $\dim V_{g,s,K}$ is expressed in terms of the modular transformation
matrices $S_{ab}$, we can  make use of some of the other algebraic properties
of conformal field theory, in particular
group-level duality \cite{006,007,008,011}, which relates $\gkn$ to $G_N(K)$,
{\it i.e.}, $SU(N)_K$ to $SU(K)_N$, etc.  The idea is that (1.1) is directly
expressible in terms of $\dim V_{g,k}$.  Using group-level duality
\cite{006,007,008,011} we then have

\be 
\lim_{N\rightarrow \infty} Z_{\S_{g}} (\gkn / \gkn ) =
\int [dA^\prime ] \d (F_A^\prime ) = {\rm Vol} (\cm^\prime )
\ee
where $A^\prime$ is the connection ($F_{A^\prime}$ the field strength) of
YM$_2$
with gauge group $\gk$.   That is \linebreak ${\displaystyle
\lim_{K\rightarrow \infty}} \gkn /\gkn$ is a BF theory associated to (YM)$_2$
at zero-area with gauge group $\gn$, while
${\displaystyle \lim_{N\rightarrow \infty}} \gkn /\gkn$ is a
{\it different} BF theory, with gauge group $\gk$.  This is an entirely
new result, made possible by the algebraic point of view.

In section 2 we will compute the volume of the moduli space of flat connections
$\cm_s$ from the Verlinde numbers, using algebraic results from conformal
field theory.  Section 3 will be devoted to a discussion of the consequences
of group-level duality for topological theories $\gkn /\gkn$.

Some relevant properties of the modular transformation matrices $S_{ab}$ are
collected in an Appendix.

\vspace{.1in}

\renewcommand{\theequation}{2.\arabic{equation}}
\setcounter{equation}{0}

\noindent{\bf II. Volume of Moduli Space of Flat Connections}

The relationship \cite{001,002,004,005} between $\gkn / \gkn$, the Verlinde
numbers $\dim V_{g,s,K}$ and BF theories allows one to calculate
Vol$(\cm_{g,s})$,
the volume of the moduli space of flat connections on the Riemann surface
$\sum_{g,s}$ of genus $g$ and $s$ punctures for gauge group $\gn$.  Witten's
\cite{001,004} approach to this question relies on Riemann-Roch theorems,
 while Blau and Thompson \cite{002} have given a path integral
argument using abelianization (see also \cite{014,015} for an argument based on
localization).  Since the Verlinde numbers \cite{005} only involve the
modular transformation matrices $S_{ab}$ of conformal field theory, once the
relationship of $\dim V_{g,s,K}$ to Vol$(\cm_{g,s})$ is established, the
remainder
of the problem may be carried out by straightforward algebraic techniques.
This latter approach was actually illustrated by Witten \cite{001} for the
special cases of $SU(2)$ and $SO(3)$. In this section we present the
algebraic computation of Vol$(\cm_{g,s})$ for the simply connected groups
$SU(N)$, $Sp(N)$, and Spin$(N)$ as well as for the non simply connected
$SO(N)$.

Equations (1.1) and (1.2) relate the topological coset theory $\gkn / \gkn$
to the BF theory of $\S_{g,s}$.  The large $K$ limit (1.1) provides
the connection to Vol($\cm_{g,s}$) since in this limit one is considering the
zero-area limit of 2-dimensional Yang--Mills theory for gauge group $\gn$.  The
Verlinde numbers for the topological correlators of primary fields $\Phi_{b}$

\be 
\langle \Phi_{b_{1}} \; \ldots \; \Phi_{b_{s}} \rangle
\ee
of $\gkn / \gkn$, where $b_i$ denotes an integrable representation of $\gkn$,
are
$$
\dim V_{g,s,K}  =  \sum_a \; S_{oa}~\!\!^{2-2g-s} \prod^s_{i=1}
S_{ab_{i}}~~~~~~~~~~~ \eqno{(2.2{\rm a})}
$$
$$
\hspace*{1in} =  S_{oo}~\!\!^{2-2g-s} \sum_a \; [\dim a]^{2-2g-s}
\sum^s_{i=1} S_{ab_{i}}
\eqno{(2.2{\rm b})}
$$
where in the sum $a$ runs over all integrable representations of $\gkn$, and
$S_{ab}$ is the modular transformation matrix.  The $q$-dimension of the
integral representation $a$ is
\renewcommand{\theequation}{2.\arabic{equation}}
\setcounter{equation}{2}
\be
[\dim a]  =  \frac{S_{oa}}{S_{oo}}
\ee
The large $K$ limit of (2.3) gives
\be 
\lim_{K\rightarrow \infty} [\dim a ] = (\dim a)
\ee
where (dim $a$) is the usual dimension of the representation $a$ of the Lie
algebra $\gn$, such that $a$ is the element on the orbit $\G (a)$ with the
smallest classical dimension, and $\G$ is the group of quantum automorphisms
of the extended Dynkin diagram of $\gkn$, which we will now examine.

There are automorphisms of the extended Dynkin diagram associated to the
affine Lie algebra $\gkn$, which we call ``classical" or ``quantum" according
to whether the automorphism in question is a symmetry of {\it only} the
usual Dynkin diagram.  Thus, automorphisms of the extended Dynkin diagrams,
which are not automorphisms of the Dynkin diagrams are quantum automorphisms.
[See Fig. 1 of Mlawer, {\it et al.} \cite{007}.]   The quantum automorphisms
generate orbits which divide the primary fields of $\gkn$ into classes.  Let
$\G$ denote the group of the quantum automorphisms.  [See Table 1 in Appendix
A.]
Then $\g  (a)$ is a new primary field obtained by the action of the element
$\g \, \e \, \G$, acting on $a$.  [There may be fixed points for some $a$.]
{}From equations (A.1) to (A.4) one can establish that for any representative
$a$ on the orbit $\G (a)$,
\be 
[\dim a] = [\dim \g (a) ] \; .
\ee
Therefore, one can rewrite (2.2) as
\be 
\dim V_{g,s,K} = S^{2-2g-s}_{oo} \sum_{\g (a)\e \G (a)}\; \sum_{[a]} \;
[\dim a]^{2-2g-s}\; \prod^s_{i=1} S_{\g (a) b_{i}}
\ee
where $[a]$ denotes a representative of the equivalence class of $a$.  If
$\g$ acted freely on every primary field $a$, then the sum on $\g (a)\:\e\:
\G (a)$ would run over the order of the center $\# Z(\hat{G}) = \# \G$
 for every primary field $a$,
where $\hat{G}$ is the universal covering group of $G$.  [See Table 1.]
However, this is not the case.  Some orbits have fixed points \cite{009}.
For fixed $N$ and fixed $G(N)$, the particular primary fields whose orbits
have fixed points depend on $K$.  For example, the adjoint representation of
$SU(3)_3$ is a fixed point for the action of $\s$, but the adjoint of $SU(3)_4$
is not.  As $K$ increases, for fixed $N$, the representations for which the
fixed points occur have increasingly larger dimension.  Thus, for
sufficiently large $K$, and surfaces with Euler characteristic $\chi (g,s) =
2-2g-s < 0$, the contribution of those representations upon which $\g (a)$
does {\it not} act freely make a negligible contribution to (2.6).  To
leading order in the large $K$ limit, they can be neglected.  Thus, our
derivation of $\dim (\cm_{g,o})$ is only valid for $g \geq 2$. [Witten
\cite{001,004} shows how to compute the divergent behavior of the sum for
$2-2g-s \geq 0$.]

The sum in (2.6) is interesting.  It produces selection rules to be satisfied
by the correlators.  Let us concentrate on those primary fields $a$ for which
$\G$ does act freely.  As discussed above, this will be adequate for the
large $K$ limit.  For these primary fields, using Appendix A we have
\newpage
$$
\sum_{\g (a)\e\G (a)} \; \prod^s_{i=1} \; S_{\g (a) b_{i}} = \sum_{\g\e\G} \;
\prod^s_{i=1} \; S_{\g (a) b_{i}}\hspace{4in}
$$
$$ 
= \# Z (SU(N)) \; \d
\left[  \sum^s_{i=1} \; r (b_i) {\rm mod}\; N \right] \; \prod^s_{i=1}
\; S_{ab_{i}} \;\;\;\;\;\;\; \mbox{for}\; SU(N)\hspace{.5in} \eqno (2.7{\rm
a})
$$
$$ 
= \# Z (Sp(N)) \; \d
\left[  \sum^s_{i=1} \; r (b_i) {\rm mod}\; 2 \right] \; \prod^s_{i=1}
\; S_{ab_{i}} \;\;\;\;\;\;\; \mbox{for}\; Sp(N)\hspace{.6in}\eqno (2.7{\rm
b})
$$
$$ 
= \# Z (\mbox{Spin}(2n+1)) \; \d
\left[  \sum^s_{i=1} \; s (b_i) {\rm mod}\; 2 \right] \; \prod^s_{i=1}
\; S_{ab_{i}} \;\;\;\;\;\;\; \mbox{for Spin} (2n + 1)\hspace{1in} \eqno
(2.7{\rm c})
$$
$$ 
= \# Z (\mbox{Spin}(2n))\; \d
\left[  \sum^s_{i=1} \; s (b_i) {\rm mod}\; 2 \right] \d
\left[ \sum^s_{i=1}  \; r(b_i) {\rm mod} \; 2 \right]
\prod^s_{i=1}\; S_{ab_{i}} \;\;\;\;\;
 \mbox{for Spin} (2n) \hspace{.4in}\eqno (2.7{\rm d})
$$
where $\# Z(\hat{G})$ is the order of the center of the universal cover of
the group $G$ as determined in Table 1, $r(b_i)$ is the number of boxes of
the Young diagram associated to the representation $b_i$ and
\renewcommand{\theequation}{2.\arabic{equation}}
\setcounter{equation}{7}
\be 
s(b_i) = \begin{array}{c}
0 \\ 1
\end{array}
\left. {\rule{0mm}{11mm}} \right\}
\;\;\;\;\;\; \mbox{ if $b_i$ is} \;\;\;\;\;\;
\left\{
\begin{array}{c}
\mbox{tensor} \\ \mbox{spinor}
\end{array}
\right\}
\end{equation}
Therefore (2.7a) denotes conservation of $N$-ality, (2.7b) and (2.7d) require
the total number of boxes in the Young diagrams associated to $b_i$ be even,
and (2.7c) and (2.7d) require the total number of spinor representations in
the correlator (2.1) be even.  For fixed points of the orbit $\g (a) \; \e \;
\G (a)$ the sum in (2.7) would still be valid if they are counted with the
multiplicity of $\# \G = \# Z (\hat{G})$.  If the requirements of (2.7) are
not satisfied, the topological correlator vanishes.  In all that follows, we
only consider non-vanishing correlators, implicitly assuming that the
selection rules (2.7) are satisfied, and therefore we will omit writing the
Kronecker deltas of (2.7).

{}From the large $K$ limit of (2.6) one obtains
\be 
\mbox{Vol}(\cm_{g,s}) = \lim_{K\rightarrow \infty}
\left[ K^{-\frac{\dim \cm_{g,s}}{2}}\; \dim \; V_{g,s,K} \right]
\ee
where
\begin{eqnarray}
\dim \; \cm_{g,s} & = & (2g + s -2)\dim \; G - s \cdot \mbox{rank}\, G
\nonumber \\
& = & (2g -2) \dim \, G + 2sR
\end{eqnarray}
with $R$ is the number of positive roots of the Lie algebra $G(N)$.  If one
normalizes group volumes as does Witten \cite{001,004} so that $U(1) = S^1$
has group volume
$2\sqrt{2} \; \pi$, then consistent with expressions of Kac and
Wakimoto \cite{010}, and Witten \cite{001,004}, one has
\be 
S_{oo}\;\; { _{\stackrel{\loon}{K\rightarrow \infty}} }
\;\; K^{-\frac{\dim G}{2}} \;
\frac
{(2\pi)^{\dim G}}
{\mbox{vol}(G)}
\ee
and
\be 
S_{ab}\;\; { _{\stackrel{\loon}{K\rightarrow \infty}} } \;\;
K^{-\frac{{\rm rank} \:G }{2}} \;
\frac{N_a (\mbox{\boldmath $\th$}_b)}{{\rm vol} (T)}(2\pi)^{{\rm rank}\, G}
\ee
where $T$ is the maximal torus of $G$ and $N_a (\mbox{\boldmath $\th$}_b)$ is
the numerator of the classical Weyl character of the
representation $a$ evaluated in the conjugacy class $\mbox{\boldmath $\th$}_b$,
 where $\mbox{\boldmath $\th$}_b$ is constant as $K \rightarrow \infty$. Notice
that the factors of ${\rm vol} (G)$ and ${\rm vol} (T)$ in (2.11) and (2.12)
are
determined by the relation of the symplectic volume of $\cm_{g,s}$ with the
scale of the Killing metric on $G$.
Therefore,\footnote{One may verify (2.11) up to an overall
numerical constant, but including the correct powers of $\pi$, by a direct
calculation, using induction.}
\be 
\prod^s_{i=1} \; S_{a b_{i}} \;\;
{ _{\stackrel{\loon}{K\rightarrow \infty}} } \;\;
K^{-\frac{s}{2}\:{\rm rank\,G}} \;
\frac{1}{({\rm vol} \: T)^{s}} \;
\prod^s_{i=1} \; N_a (\mbox{\boldmath $\th$}_{b_{i}})\;
(2\pi)^{s\: {\rm rank}\, G} .
\ee
Combining (2.6) with with (2.11)--(2.13) we obtain (for $2-2g-s < 0)$
\begin{eqnarray}
\lim_{K\rightarrow \infty} \dim \; V_{g,s,K} & = & \left(\frac{K}{4\pi^2}
\right)
^\frac{\dim \, \cm_{g,s}}{2} \;
\frac{(\mbox{vol} \: G)^{2g+s-2}}{(\mbox{vol} \: T)^s}
\nonumber \\
& \cdot & \# Z(\hat{G}) \; \sum_a \; (\dim \: a)^{2-2g-s} \;
\prod^s_{i=1} \; N_a (\mbox{\boldmath $\th$}_{b_{i}})
\end{eqnarray}
where $a$ is the ``smallest"
representation on the orbit $\G (a)$.  Therefore, from (2.9) and (2.14) we
arrive at
\begin{eqnarray}
\mbox{Vol} (\cm_{g,s}(G)) & = &
\frac{\# Z(\hat{G})}
{(2\pi )^{\dim \, \cm_{g,s}} }
\;
\frac{(\mbox{vol} \: G)^{2g+s-2}}
{(\mbox{vol} \: T)^s}
\nonumber \\
& \cdot & \sum_a \: (\dim a)^{2-2g-s} \; \prod^s_{i=1} \; N_a
(\mbox{\boldmath $\th$}_{b_{i}})
\end{eqnarray}
which agrees with the result of Witten \cite{001,004}.

In order to consider $SO(N)$ rather than Spin$(N)$, one follows the procedure
of Witten  \cite{001,004} detailed for $SO(3)$, and the methods above, to
find that (2.15) remains valid for $SO(N)$, {\it except} that the sum
on $a$, and the insertions {\it only} involve tensor representations.
In the BF theory this corresponds to summing over all topologies of
$SO(N)$-bundles on $\Sigma_{g}$ \cite{004}.

Thus, appealing to algebraic information has given an alternative derivation of
(2.15).  In the next section the algebraic approach allows us to make use of
group-level duality, which leads to new results; not just a dfferent
derivation of known ones.

\vspace{.1in}

\noindent{\bf III. Group-level Duality}

Group-level duality \cite{006,007,008,011} has provided a number of interesting
insights into conformal and topological field theories.  In this section we
consider the application of this duality to $\gkn / \gkn$ topological
theories, and show that $\displaystyle{\lim_{N\rightarrow \infty}} \gkn
/\gkn$ is a BF theory with gauge group $\gk$ {\it if} $\gn$ is a
classical group. That is, the limit yields a zero-area two-dimensional
Yang-Mills theory with gauge group $G(K)$, and therefore computes
Vol$\cm_{g,s}(\gk )$. Notice that although there is a path integral argument
for
showing that the large $K$ limit of $\gkn/\gkn$ is a ($G(N)$) BF theory, no
such argument exists (at present) for the large $N$ limit. There are
difficulties with the group-level duality
for Spin$(N)$, which prevents a similar conclusion for this case.  A number
of properties of the modular transformation matrices are summarized in Eqn's
(A.6) to (A.13).  Further details can be found in \cite{007}.
\newpage

\noindent{\boldmath $Sp(N)_{\rm K}$}

The simplest application of the group-level duality \cite{007} is for
$Sp(N)$,
where $N$ is the rank of the group. In this case there is a 1--1
correspondence between primary fields of $Sp(N)_K$ and of $Sp(K)_N$, by
means of the map from a Young diagram for representation $a$ to the diagram
which interchanges rows and columns, corresponding to the representation
$\tilde{a}$.  From (A.7) and (2.2) we therefore have
\renewcommand{\theequation}{3.\arabic{equation}}
\setcounter{equation}{0}
\be 
\dim V_{g,s} (Sp(N)_K) = \dim V_{g,s} (Sp(K)_N ) \; .
\ee
Then from (2.11) and (2.12) we obtain
\be
{\rm Vol}\; \cm_{g,s} (Sp(K)) =
\lim_{N\rightarrow \infty}
\left[ \; N^{-\frac{\dim \cm_{g,s} (Sp(K))}{2}}
\dim V_{g,s} (Sp(N)_K)\; \right]
\ee
and for the correlators (see also ref. \cite{008})
$$
\langle\Phi_{b_{1}}\cdots\Phi_{b_{s}}\rangle_{Sp(N)_{K}} =
\langle\Phi_{\tilde{b}_{1}}\cdots\Phi_{\tilde{b}_{s}}\rangle_{Sp(K)_{N}}
$$

\noindent{\boldmath $SU(N)_K$}

In this case, group-level duality \cite{006,007,011} is not a mapping between
primary fields, but only between cominimal orbits, whose elements are denoted
by $\s$ in Table 1, and Mlawer, {\it et al.} \cite{007}.  From (A.6) and
(2.2) we find
\be
\dim V_{g,s} (SU(N)_K) = \left( \frac{N}{K} \right)^g \dim V_{g,s}
(SU(K)_N) \; .
\ee
The actual correlators satisfy \cite{007,008}
\be
\langle \Phi_{b_{1}} \cdots \Phi_{b_{s}} \rangle_{SU(N)_K}
= \left( \frac{N}{K} \right)^g
\langle \Phi_{\s^{p_{1}} (\tilde{b}_1)} \cdots
\Phi_{\s^{p_{s}} (\tilde{b}_s)}
\rangle_{SU(K)_N}
\ee
for any set of integers $p_1 \cdots , p_s$ such that
$\sum^s_{i=1} p_i = - \D \; \mbox{mod} K$, with $\D = \sum^s_{j=1} r
(b_j)/N$, where $r (b_j)$ denotes the number of boxes in the tableau for
representation $b_j$.  Here $\s^p (b_i)$ denotes the $p^{\rm th}$
representative in the cominimal orbit.  Hence we find
\be 
\lim_{N\rightarrow \infty}
\left[ N^{-\frac{2g+\dim \cm_{g,s} (SU(K))}{2}}
\dim V_{g,s} (SU(N)_K) \right] = K^{-g} \; \mbox{Vol}\; \cm_{g,s} (SU(K))
\ee

\noindent{\boldmath $SO(N)_K$}

As discussed in the next-to-last paragraph of Section 2, we are restricted to
tensor representations only.  The dual map is not between primary fields, but
between cominimal orbits of $\s$, which have length 2 for both $SO(N)_K$ and
$SO(K)_N$.  From (2.2a) (A.9) and (A.13) one easily obtains
\be 
\dim V_{g,s} (SO(N)_K ) = \dim V_{g,s} (SO(K)_N) \; .
\ee
Thus
\be 
\lim_{N\rightarrow \infty}
\left\{ N^{\frac{-\dim \cm_{g,s} (SO(K))}{2}}
\dim V_{g,s} (SO(N)_K) \right\}
 = \mbox{Vol}\; \cm_{g,s} (SO(K)) \; .
\ee
The cominimally reduced correlators satisfy
\be 
\langle \Phi_{b_{1}} \cdots \Phi_{b_{s}} \rangle _{SO(N)_K} =
\langle \Phi_{\tilde{b}_1} \cdots \Phi_{\tilde{b}_s} \rangle _{SO(K)_N}\; .
\ee
Notice that for $SO(even)_{odd}$ duality, one must be careful with the $\rho$
symmetry, which in this case maps tensor representations to spinor
representations (see the Appendix A). Omitting this symmetry takes care of the
factor of 2 in $\# Z({\rm Spin} (2n))/\# Z({\rm Spin} (2k+1))$.
Thus all the classical groups exhibit the group-level duality (3.1), (3.3),
and (3.6) for the Verlinde numbers and (3.2), (3.5) and (3.7) for the
Vol$(\cm_{g,s})$.

\noindent{\bf Spin{\boldmath$(N)_K$}}

The situation for Spin$(N)_K$/Spin$(N)_K$ is very different, as the
group-level duality for $S_{ab}$ is quite complicated.  [See (A.8) {\it ff}
and Mlawer, {\it et al.} \cite{007}.]  For example, for Spin$(2n)_{\rm 2k}$
and Spin$(2n)_{2k+1}$ one does not even know if there is a dual map for
$S_{ab}$ involving spinor representations.  Equations (A.8) and
(A.10)--(A.12) are the dual maps for spinors for Spin$(2n+1)_{2k+1}$.
Therefore this is the only case we will discuss, but with no results
comparable to those obtained for the classical groups.

\noindent{\bf Spin\boldmath$(2n+1)_{2k+1}$}

Let us consider the Verlinde dimensions for a surface of genus $g$, without
insertions, which is sufficient to illustrate the difficulties with
group-level duality for this case.  We can write
$$
\dim V_{g,o} ({\rm Spin} (2n+1)_{2k+1})\hspace{2.3in}
$$
$$
= S^{2-2g}_{oo} \; \sum_a \; \left( \frac{S_{oa}}{S_{oo}} \right)^{2-2g}
\hspace{1.6in}
\eqno{(3.9{\rm a})}
$$
$$
= (S_{oo})^{2-2g} \;
\left\{ \sum_t \; \left( \frac{S_{ot}}{S_{oo}} \right)^{2-2g} +
\sum_s \;
\left( \frac{S_{os}}{S_{oo}} \right)^{2-2g} \right\}
\eqno{(3.9{\rm b})}
$$
where in (3.9b) we divided the sum into tensor and spinor contributions.
However, in order to use (A.8) and (A.10)--(A.12) for the duality, we must
separate the spinor representations into two classes; those $s=b$, for which
$\ell_1 (b) = K/2$, and those $s=b^\prime$, for which $\ell_1 (b^\prime) <
K/2$, where $\ell_1$ is the length of the first row of the Young tableaux.
Then from \cite{007,012} (A.8) to (A.12) we can show that
\renewcommand{\theequation}{3.\arabic{equation}}
\setcounter{equation}{9}
\begin{eqnarray}
\lefteqn{\dim V_{g,o} (\mbox{Spin} \;(2n+1)_{2k+1})} \nonumber \\[.1in]
& = & (\tilde{S}_{oo})^{2-2g} \left\{ \sum_{\tilde{t}} \;
\left( \frac{\tilde{S}_{o\tilde{t}}}{\tilde{S}_{oo}} \right) ^{2-2g} \right.
  \nonumber\\[.1in]
& + & 2^{g} \; \sum_{\hat{b}^\prime} \;
\left( \frac{\tilde{S}_{o\hat{b}^\prime}}{\tilde{S}_{oo}} \right)^{2-2g}
\nonumber \\[.1in]
& + & 2^{-g} \; \sum_{\hat{b}} \; \left.
\left( \frac{\tilde{S}_{o\hat{b}}}{\tilde{S}_{oo}} \right)^{2-2g}
\right\} \; .
\end{eqnarray}
Hence
\be 
\dim V_{g,o} ({\rm Spin}\;(2n+1)_{2k+1} ) =
\dim V_{g,o} ({\rm Spin}\;(2k+1)_{2n+1} ) \;\;\;
\mbox{for $g=0$ only.}
\ee

Therefore, we should restrict our attention to genus zero surfaces.  Consider
the special case where all insertions are tensor representations.  Then
proceeding as in (3.9)--(3.10) we obtain

\begin{eqnarray}
\lefteqn{\dim V_{o,s} (\mbox{Spin}(2n+1)_{2k+1})} \nonumber \\[.1in]
& = & (S_{oo})^2 \left\{ \sum_a \;
\left( \frac{S_{oa}}{S_{oo}} \right) ^{2-s}\;
\prod^s_{i=1} \; \left(
\frac{S_{ad_{i}}}{S_{oo}} \right) \right\}
\nonumber\\[.1in]
& = & (\tilde{S}_{oo})^2 \left\{ \sum_{\tilde{t}}\;
\left( \frac{\tilde{S}_{o\tilde{t}}}{\tilde{S}_{oo}} \right) ^{2-s} \; \right.
\prod^s_{i=1} \;
\left( \frac{\tilde{S}_{\tilde{t}\tilde{d}_{i}}}{\tilde{S}_{oo}} \right)
\nonumber\\[.1in]
& ~~~~+ & \sum_{\hat{b}^\prime}\;
\left( \frac{\tilde{S}_{o\hat{b}^{\prime}}} {\tilde{S}_{oo}} \right) ^{2-s}
\; \prod^s_{i=1} \; (-1)^{\S_{i}r(d_{i})} \left(
\frac{\tilde{S}_{\hat{b}^{\prime}\tilde{d}_{i}}}
{\tilde{S}_{oo}} \right)\nonumber\\[.1in]
& ~~~~+ & \sum_{\hat{b}}\;
\left. \left( \frac{\tilde{S}_{o\hat{b}}} {\tilde{S}_{oo}} \right) ^{2-s}
\; \prod^s_{i=1} \; (-1)^{\S_{i}r(d_{i})} \left(
\frac{\tilde{S}_{\hat{b}\tilde{d}_{i}}}{\tilde{S}_{oo}} \right) \right\}
\end{eqnarray}
Thus, for $s$ tensor insertions in a genus zero surface,
\be
\dim \: V_{o,s} ({\rm Spin}(2n+1)_{2k+1}) \; = \;
\dim \: V_{o,s} ({\rm Spin}(2k+1)_{2n+1})
\ee
{\it if} the following selection rules hold:\\
a) ~If $\sum_i r(d_i) = 2 \mbox{\boldmath $Z$}$, then the topological
amplitude for $s$ tensor insertions $\{ d_i \}$ are dual to that for $s$
tensor insertions $\{ \tilde{d}_i \}$ where $d_i$ is mapped to $\tilde{d}_i$.
\\
b) ~If $\sum_i \, r(d_i) = 2 \mbox{\boldmath$Z$} +1$, then since
$S_{\s (d) b} = - S_{db}$ for $b$ spinor, one should map an odd number of the
$d_i$ {\it not} to $\tilde{d}_i$, but rather to $\s (\tilde{d}_i )$.

A computation similar to (3.9)--(3.10) shows that we cannot establish a
group-level duality of Verlinde dimensions for one or more spinor insertions.
Further, we cannot take the large $k$ or large $n$ limit of (3.11) because of
the problem with fixed points of the orbits of $\G$, discussed in (2.6)~{\it
ff}, since the summations will diverge.  However, for 3 or more tensor
insertions and genus zero, (3.12) and (3.13) imply a result analogous to (3.7)
 {\it i.e.},
\begin{eqnarray}
\lim_{n\rightarrow\infty} \dim \: V_{o,s} ({\rm Spin} (2n+1)_{2k+1})
& = & {\rm Vol} \: \cm_{o,s} ({\rm Spin} (2k+1)) \\
& ~ &  \mbox{for} \; s \geq 3 \;
\mbox{tensor insertions.} \nonumber
\end{eqnarray}

We have observed that group-level duality has only a limited scope for
Spin$(2n+1)_{2k+1}$, and with no results at all available for the other
Spin$(N)_K$ cases.  Our difficulties with group-level duality
for Spin$(N)_K$ have been demonstrated using only algebraic methods.  It
would be interesting if a more geometric understanding of the ``obstruction"
to group-level duality for Spin$(N)_K$ could be found.

Witten \cite{011} relates the Verlinde formulae, and the gauged WZW model of
$G_K/G_K$ to the quantum cohomology of the Grassmannian manifold, $G_V(K,N)$
where $G_V(K,N)$ is the space of all $K$ dimensional subspaces of a fixed
complex vector space $V \simeq  \mbox{\boldmath$C$}^N$.  As a result of
group-level duality of the Verlinde algebra of $U(K)$ at level $(N-K,N)$ with
that of $U(N-K)$ at level $(K,N)$, one has the isomorphism of the quantum
cohomology rings for $V$ and its dual space $V^*$, {\it i.e.},
\be
G_V(K,N) \cong G_{V^*} (N-K,N) \; .
\ee
This cohomology ring is closely related to the chiral ring of the $N=2$
superconformal field theory often described by a $U(N)/U(K) \times U(N-K)$
coset model.

One may consider similar issues for symplectic and real Grassmannians.  For $V
\simeq \mbox{\boldmath$H$}^N$ \\
($N$-dimensional symplectic space), one has the
group-level duality of $Sp(N)_K$ and $Sp(K)_N$,
which will lead to the isomorphism (3.15).  Similarly for
$V \simeq \mbox{\boldmath$R$}^N$, the
group-level duality of the $SO(N)_K$ and $SO(K)_N$; ({\it i.e.}, tensor
representations only) also gives the isomorphism (3.15).  However, recall
that group-level duality does {\it not} hold in general between the
universal covering groups Spin$(N)_K$ and Spin$(K)_N$.
Since Spin$(N)/({\rm Spin}(K) \times {\rm Spin}(N-K) / Z_2)$ is a real
Grassmannian it is not clear to us what implications this has for quantum
cohomology rings.  \\

\noindent{\bf Summary}

In this paper we have illustrated the usefulness of the algebraic approach to
the computation of
Vol(${\cal M}$), the volume of the moduli space of flat connections of the
gauge group $G(N)$ on the Riemann surface $\Sigma_{g,s}$. As a consequence,
we were able to use group-level duality to show that
$\displaystyle \lim_{N\rightarrow \infty} \gkn / \gkn$ provides a computation
of Vol(${\cal M}^\prime$), the dimension of the moduli space of flat
connections for the group $G(K)$, if $G(N)$ is a classical group. This is
a new result which
hasn't been obtained by path-integral arguments. This raises the question
whether this same result can be obtained by functional-integral methods.
 We discussed the algebraic
difficulties encountered in applying group-level duality to ${\rm Spin}(N)_K$.
One would like to have a geometric understanding of why group-level duality
fails for this case.

{\bf Acknowledgements:}
We would like to thank Prof. A. V. Ramallo for helpful comments on the
manuscript.

\newpage

\renewcommand{\theequation}{A.\arabic{equation}}
\setcounter{equation}{0}

\noindent{\bf Appendix A}

We need to consider certain automorphisms of the extended Dynkin diagrams
associated to the classical affine Lie algebras, and the transformations of
$S_{ab}$ induced by these automorphisms.  Properties of $S_{ab}$ under
group-level duality are also required in the body of the text.  All the
relevant material is to be found in Mlawer, {\it et al.} \cite{007}, so that
we only give a brief summary of the necessary results.  The reader should
refer to \cite{007} for further details.

\noindent{\bf 1. Extended Dynkin Diagrams}

We are interested in those automorphisms of the extended Dynkin diagrams
which are {\it not} automorphisms of the Dynkin diagrams of the
classical, semisimple Lie algebras.  Call these the quantum automorphisms.
{}From Fig. 1
of ref. \cite{007}, we record the following quantum automorphisms:\\
1)~~ $SU(N)$: ~an automorphism $\s$, of order $N$, which generates the $Z_N$
orbit of cominimal equivalence.\\
2)~ $Sp(N)$: ~an automorphism $\r$, with $Z_2$ symmetry, which complements
the associated Young diagram of a representation.\\
3)~ ${\rm Spin}(2n+1)$: ~$\s$ is a $Z_2$ symmetry which generates cominimal
equivalence classes.\\
4)~ ${\rm Spin}(2n)$: ~$\s$ is a $Z_2$ symmetry which generates cominimal
equivalence
classes and $\r$ is a $Z_2$ symmetry which complements associated Young
diagrams of a representation.

It should be noted that the order of the center of the universal
covering groups of the classical groups is identical to
the order of the quantum symmetries of the extended Dynkin diagrams, as
summarized in the Table I.

\begin{center}
\begin{tabular}{c|c|c|c}
          &                 & \# of elements of & \\[-.2in]
$\hat{G}$ & $\# Z(\hat{G})$ & quantum orbits & generators \\
\hline
$SU(N)$ & N & N & $\s$ \\
$Sp(N)$ & 2 & 2 & $\r$ \\
Spin$(2n)$ & 4 & $2\cdot 2$ = 4 & $\s,\r$\\
Spin$(2n+1)$ & 2 & 2 & $\s$
\end{tabular}
\end{center}

\vspace{.1in}

\noindent{\bf 2. Quantum Automorphisms and $\mbox{\boldmath$S_{ab}$}$}

\noindent 1) $SU(N)_K$:
\be 
S_{\s (a)b} = e^{-2\pi ir(b)/N} S_{ab}
\ee
where $r(b)$ is the number of boxes of the Young diagram representing the
primary field of $SU(N)_K$.\\
2) $Sp(N)_K$:
\be
S_{\r (a)b} = e^{i\pi r(b)} S_{ab}
\ee
3) ${\rm Spin}(N)_K$:
\be
S_{\s (a)b} = \left\{ \begin{array}{ll}
S_{\e (a)b} & \mbox{for $b$ a tensor representation}\\
-S_{\e (a)b} & \mbox{for $b$ a spinor representation}
\end{array}  \right.
\ee
where $\e (a) = a$ for all representations of ${\rm Spin}(2n+1)_K$, and for all
non-self-associate representations of ${\rm Spin}(2n)_K$. See \cite{007} for
details appropriate to self-associate representations.

For ${\rm Spin}(2n)_K$ we also need the automorphism $\r$.  For odd $K, \;\r$
exchances tensor and spinor representations.  For even $K, \; \r$ maps
tensors to tensors and spinors to spinors.  Then for ${\rm Spin}(2n)_K$
\be
S_{\r (a)b} =  e^{i\pi r(b)} S^*_{ab} =
\left\{ \begin{array}{ll}
 e^{i\pi r(b)} S_{ab}  & : ~~~ n \; \mbox{even}\\
 e^{i\pi r(b)} S_{\e (a)b}  & : ~~~ n \; \mbox{odd}
\end{array}  \right.
\ee

Recall that the quantum dimensions of the integrable representation $a$ is
\be
[a] = \frac{S_{oa}}{S_{oo}}
\ee
Therefore, the automorphisms (A.1) to (A.4) preserve quantum dimensions.

{\bf 3. Group-Level Duality and $\mbox{\boldmath $S_{ab}$}$}

\underline{$SU(N)_K$}: ~Associate $a$ with its reduced diagram, with $r(a)$
boxes.  Then $\tilde{a}$ denotes the representation of $SU(K)_N$ of the
transpose of this reduced diagram.  Then \cite{011,007}
\be
S_{ab} = \sqrt{\frac{K}{N}} \; \exp [-2\pi i r(a) r(b) / NK ]
\tilde{S}^*_{\tilde{a}\tilde{b}}
\ee
\underline{$Sp(N)_K$}:
\be
S_{ab} = \tilde{S}_{\tilde{a}\tilde{b}}
\ee

The situation for ${\rm Spin}(N)_K$ is quite complicated, and perhaps
incomplete.
See Mlawer, {\it et al.} \cite{007} for information to supplement the results
summarized below.

\noindent\underline{Spin$(2n+1)_{2k+1}$}:
\be
S_{Ab} = (-1)^{r(A)} \: s^{t(b)-1/2} \; \tilde{S}_{\tilde{A}\hat{b}}
\ee
where
$A$ is a cominimally reduced tensor representation,
$b$ is a cominimally reduced spinor representation and
$t(b) = 1$ if $\ell_1 (A) = \frac{1}{2} K$ and 0 otherwise. For $A$ and $B$
tensor representations one has,
\be
S_{AB} = \tilde{S}_{\tilde{A}\tilde{B}}
\ee
In fact this result holds even if the tensor representations are not
cominimally reduced.  The duality of the spinor-spinor type is to be found in
Fuchs and Schweigert \cite{013}, eqn's (3.20) and (3.30) {\it ff}.
\newpage
\be
S_{bb^\prime} \;\; = \;\; \tilde{S}_{\hat{b}\hat{b}^\prime}\;\; = \;\; 0 \\
\ee
$$
\mbox{if}\;\;\; \left\{
\begin{array}{lcl}
b & \mbox{has} & \ell_1(b) \;\; = \;\; K/2\\
b^\prime & \mbox{has} & \ell_1(b^\prime)\;\; < \;\; K/2
\end{array} \right.
$$
and \\
\be
S_{bb^\prime}\;\;  = \; 0 \; ; \;\;\; \tilde{S}_{\hat{b}\hat{b}^\prime} \;\;
\neq \;\; 0
\ee
$$
\mbox{if} \;\;\;\ell_1(b) \;\; = \;\; \ell_1(b^\prime) \;\; = \;\;
\frac{1}{2} \: K\;.
$$
$$
\mbox{[since} \;\;\; \ell_1 (\hat{b})\; < \; N/2 \; , \;\;\;
\ell_1(\hat{b}^\prime)\; < \; N/2 ].
$$
Similarly,
\be
S_{bb^\prime}\;\; \neq \;\; 0 \;\; ; \;\;\;\; \tilde{S}_{\hat{b}\hat{b}^\prime}
\;\; = \;\;0
\ee
$$
\mbox{if} \;\;  \ell (b)\; < \; \frac{1}{2} \: K \; ; \;\;\; \ell_1 (b^\prime
) \; < \; \frac{1}{2} \: K
$$
where $\ell_1 (b)$ is the first row length of the Young tableau associated to
$b$, which is half-integeral for a spinor representation.  The dual map
$b\rightarrow \hat{b}$ is defined in the paragraph following (A.22) of
Mlawer, {\it et al.} \cite{007} for spinor representations.

\noindent\underline{Spin$(2n)_{2k}$ and Spin$(2n)_{2k+1}$}

\be
S_{AB} = \tilde{S}_{\tilde{A}\tilde{B}}
\ee
for $A$ and $B$ tensor representations, independent of whether they are
cominimally reduced or not.  Results for this case are not known if spinor
representations are included.

\end{document}